\pdfoutput=1
\documentclass[9pt,twocolumn,twoside]{pnas-new}
\templatetype{pnasresearcharticle} 
\title{Rotational sound in disordered granular materials}
\author[a,b,1]{Kuniyasu Saitoh}
\author[c]{Rohit K. Shrivastava}
\author[c]{Stefan Luding}
\affil[a]{Research Alliance Center for Mathematical Sciences, Tohoku University, 2-1-1 Katahira, Aoba-ku, Sendai 980-8577, Japan}
\affil[b]{WPI-Advanced Institute for Materials Research, Tohoku University, 2-1-1 Katahira, Aoba-ku, Sendai 980-8577, Japan}
\affil[c]{Faculty of Engineering Technology, MESA+, University of Twente, Drienerlolaan 5, 7522 NB, Enschede, The Netherlands}
\leadauthor{Saitoh}
\significancestatement{
Elastic mechanical waves in granular media have a wide range of application for technology and pose many fundamental challenges for research.
The microstructure of granular material is strongly disordered and friction acts between particles.
The former affects the modes of momentum and energy transport, while the later induces micropolar rotations of particles.
The rotations propagate through granular media as rotational modes and much less attention has been paid to the rotational sound in disordered systems.
Here, we show that the rotational modes exhibit a transition from optical-like fast oscillations to the acoustic behavior for short wave lengths.
Our findings suggest that frequency-bands of the rotational modes can be drastically altered by introducing configurational disorder in the microstructure of granular materials.
}
\authorcontributions{Author contributions: K.S. and S.L. designed research; K.S. performed research; and K.S., R.K.S., and S.L. wrote the paper.}
\authordeclaration{The authors declare no conflict of interest.}
\correspondingauthor{\textsuperscript{2}To whom correspondence should be addressed. E-mail: kuniyasu.saitoh.c6@tohoku.ac.jp}
\begin{abstract}
We employ numerical simulations to understand the evolution of elastic standing waves in disordered frictional disk systems, where the dispersion relations of rotational sound modes are analyzed in detail.
As in the case of frictional particles on a lattice, the rotational modes exhibit an ``optical-like'' dispersion relation in the high frequency regime,
representing a shoulder of the vibrational density of states and fast oscillations of the autocorrelations of rotational velocities.
A lattice-based model describes the dispersion relations of the rotational modes for small wave numbers.
The rotational modes are perfectly explained by the model if tangential elastic forces between the disks in contact are large enough.
If the tangential forces are comparable with or smaller than normal forces, the model fails for short wave lengths.
However, the dispersion relation of the rotational modes then follows the model prediction for transverse modes, implying that the fast oscillations of disks' rotations switch to acoustic sound behavior.
We evidence such a transition of the rotational modes by analyzing the eigen vectors of disordered frictional disks and identify upper and lower limits of the frequency-bands.
We find that those are not reversed over the whole range of tangential stiffness as a remarkable difference from the rotational sound in frictional particles on a lattice.
\end{abstract}
\dates{This manuscript was compiled on \today}
\doi{\url{www.pnas.org/cgi/doi/10.1073/pnas.XXXXXXXXXX}}
\begin{document}
\maketitle
\thispagestyle{firststyle}
\ifthenelse{\boolean{shortarticle}}{\ifthenelse{\boolean{singlecolumn}}{\abscontentformatted}{\abscontent}}{}
\dropcap{G}ranular material exists at all spatial scales in nature, whether it is soil or space dust.
An important feature of such media is its momentum and energy transport characteristics,
either for oil/gas exploration, geotechnical investigations of soil, or understanding of seismic waves and earthquakes \cite{seismic_wave}.
Sound waves in granular material are also useful to determine its mechanical properties but, to better interpret measurements, a model which incorporates the microstructure is required \cite{psheng}.
In addition, tangential forces (and friction) between grains in contacts are intrinsic to granular material, which induce micropolar rotations of the constituents.
Incorporating the rotational degrees of freedom for the internal microstructure of granular material has been the basis for Cosserat continuum theory \cite{cosserat}.
So far, one of the most striking aspects of Cosserat behavior (micropolar rotations of constituent grains) is the existence of \emph{rotational sound} \cite{rot_mode0}.
This has been extensively studied by experiments \cite{rot_mode2} and numerical simulations \cite{rot_mode3,stefan-N4a,stefan-N4b} of frictional granular crystals,
where the characteristic dispersion relations, beyond Cosserat, are well predicted by the theory of frictional particles on a lattice \cite{rot_mode1,stefan-N6}.

Despite these successes of continuum theory, spatial configurations of the constituent grains in nature are mostly disordered \cite{stefan-N2,stefan-N3},
where small deviations from the lattice \cite{stefan-N1} display the characteristic low pass behavior.
Recently, various anomalies in acoustic sound in disordered media have been clarified by experiments on amorphous solids \cite{sound4,sound5}
and numerical simulations of randomly arranged particles \cite{sound0,sound1,sound2,sound3},
where small dips in phase speeds and deviations from the theory of Rayleigh scattering are commonly observed.
Anomalies in the vibrational density of states (vDOS),\ e.g.\ the boson peak near the glass transition temperature
\cite{boson0,boson1,boson2,boson3,boson4,stefan-N5a,stefan-N5b}
and a characteristic plateau near the jamming transition \cite{vm0,vm1,vm2,vm3}, are not predicted by the classical theory of elasticity.
Moreover, shock waves \cite{shock0,shock1,shock2,shock5} and solitary wave propagation \cite{shock3,shock4} are interesting properties of real disordered systems
though they are due to nonlinearity of the interaction forces and anharmonic vibrations.
Neglecting the latter phenomena and reducing on a rather simple two-dimensional model systems,
it is feasible to focus on the question: How does configurational disorder alter the dispersion relations of rotational sound?

In this paper, we investigate sound in disordered frictional disks by numerical simulations.
Introducing the \emph{dynamical matrix}, we analyze effects of tangential forces on the vDOS and demonstrate the evolution of purely elastic standing waves of \emph{longitudinal} (L), \emph{transverse} (TR), and \emph{rotational} (RT) \emph{modes}.
We calculate autocorrelation functions and spectra of L, TR, and RT velocities to find their dispersion relations.
Then, we quantify the dependence of the dispersion relations on the strength of the tangential forces,
where we introduce a modified lattice-based model to describe the ``optical-like'' branch of the rotational sound modes.
We show that the RT mode switches from optical-like to acoustic branches at characteristic wave lengths, which monotonously increase with decreasing the strength of tangential forces.
We examine closer these transitions by analyzing the eigen vectors of the disk system and also discuss the frequency-bands of the RT mode.

\section*{Results}
To study sound in disordered frictional disks, we introduce their dynamical matrix.
If the system consisting of $N$ disks is initially in mechanical equilibrium,
small vibrations of the disks around initial positions, $\{\mathbf{r}_i(0)\}$ ($i=1,\dots,N$), are described by the equations of motion,
\begin{equation}
\mathcal{M}\ddot{\mathbf{u}}(t) = -\mathcal{D}\mathbf{u}(t)~,
\label{eq:motion}
\end{equation}
where $t$ denotes time and the $3N$-dimensional displacement vector, $\mathbf{u}(t)\equiv(\mathbf{u}_1(t),\theta_1(t),\dots,\mathbf{u}_N(t),\theta_N(t))^\mathrm{T}$,
includes translational displacements in $xy$-plane, $\mathbf{u}_i(t)\equiv\mathbf{r}_i(t)-\mathbf{r}_i(0)$, and angular displacements, $\theta_i(t)$.
On the left-hand-side of Eq.\ [\ref{eq:motion}], the $3N\times3N$ \emph{mass matrix}, $\mathcal{M}$, is diagonal (Eq.\ [\ref{eq:mm}]).
On the other hand, $\mathcal{D}$ is the $3N\times3N$ dynamical matrix which consists of second derivatives of elastic energy
(see \textbf{Supplementary Information} (SI)).
In this study, we model the elastic energy by harmonic potentials in normal and tangential directions (Eq.\ [\ref{eq:eij}]), where the normal (tangential) stiffness is $k_n$ ($k_t$).
For the sake of simplicity, we assume that every disk has the same mass, $m$, and prepare initial disordered configurations, $\{\mathbf{r}_i(0)\}$, by molecular dynamics (MD) simulations (see \textbf{Materials and Methods}).

\subsection*{Vibrational density of states}
\begin{figure}
\centering
\includegraphics[width=.8\linewidth]{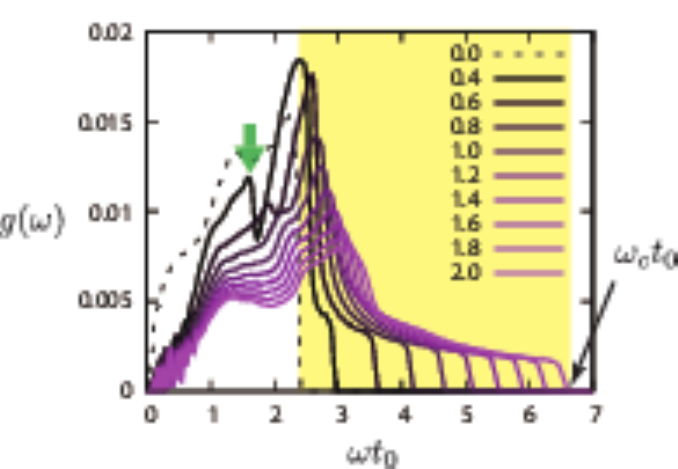}
\caption{
The vDOS of disordered frictional disks (the solid lines), where the area fraction is $\phi=0.9$ and the stiffness ratio, $\rho$, increases as listed in the legend.
The dotted line is the vDOS of disordered frictionless disks,\ i.e.\ $\rho=0$, with the same area fraction.
The shaded region represents the high frequency-band for $\rho=2$, where the vDOS ends at cutoff frequency, $\omega_c$, as indicated by the arrow.
The green vertical arrow indicates a secondary peak of vDOS for $\rho=0.4$.}
\label{fig:vdof}
\end{figure}

Assuming vibrational motions of the disks around initial positions, we substitute $\mathbf{u}(t)=\bar{\mathbf{u}}e^{I\omega t}$ into Eq.\ [\ref{eq:motion}],
where $\bar{\mathbf{u}}$, $I$, and $\omega$ are the amplitude, imaginary unit, and angular frequency, respectively.
Then, we numerically solve an eigenvalue problem, $\mathcal{M}^{-1}\mathcal{D}\bar{\mathbf{u}}_q=\omega_q^2\bar{\mathbf{u}}_q$,
to find the eigen frequencies, $\omega_q$, and eigen vectors, $\bar{\mathbf{u}}_q$ ($q=1,\dots,3N$).
Figure \ref{fig:vdof} displays distributions of the eigen frequencies,\ i.e.\ vDOS of disordered frictional disks (solid lines), where we averaged every vDOS over $100$ different initial positions of $N=2048$ disks.
Here, we change the stiffness ratio, $\rho\equiv k_t/k_n$ (as listed in the legend), and also plot the result of frictionless disks,\ i.e.\ $\rho=0$ (dotted line).
The highest peak around $\omega t_0\simeq2.3$ for frictionless disks (with the time unit, $t_0\equiv\sqrt{m/k_n}$) shifts to higher frequencies with increasing the stiffness ratio.
In addition, a shoulder, which does not exist in the vDOS of frictionless disks, develops for higher frequencies ($2.3\lesssim\omega t_0\lesssim6.6$)
so that the \emph{cutoff frequency}, $\omega_c$, at the right end of vDOS,\ i.e.\ $g(\omega>\omega_c)=0$ (indicated by the arrow in Fig.\ \ref{fig:vdof}), greatly increases from $2.3t_0^{-1}$ to $6.6t_0^{-1}$.
Therefore, the high frequency-band (the shaded region in Fig.\ \ref{fig:vdof}) is characteristic of disordered frictional disks.
Note that there is a secondary peak if the stiffness ratio is small (as indicated by the vertical arrow in Fig.\ \ref{fig:vdof})
and the overpopulation of low frequency states for $\rho=0$ vanishes and becomes linear within the fluctuations for $\rho>0$.
Because the vDOS in the high frequency regime is insensitive to the area fraction of the disks, $\phi$ (see Fig.\ S1), we use $\phi=0.9$ in the following analyses.
%

\subsection*{Dispersion relations}
To investigate how elastic waves evolve in disordered frictional disk systems, we employ a similar method as Gelin \emph{et al.} \cite{sound0}:
We numerically integrate the equations of motion,\ Eq.\ [\ref{eq:motion}], under periodic boundary conditions.
The number of disks is now increased to $N=32768$ and initial velocities of the disks are given by sinusoidal standing waves,
\begin{equation}
\{\dot{\mathbf{u}}_i(0),\dot{\theta}_i(0)\}=\{\mathbf{A},A_\theta\}\sin(\mathbf{k}\cdot\mathbf{r}_i(0))~,
\label{eq:initial}
\end{equation}
where $\mathbf{A}$ and $A_\theta$ are small amplitudes and $\mathbf{k}$ is the wave vector.
Combining different amplitudes and wave vectors, we activate three different elastic waves,\ i.e.\ L, TR, and RT modes, as listed in Table \ref{tab:modes}
\footnote{Due to the tangential force and \emph{interlocking} of the disks, every mode is excited by any combinations of the amplitudes and wave vectors.
For example, the combination in the first line of Table \ref{tab:modes} excites not only the L mode, but also the TR and RT modes though their intensities are quite weak.}.
The RT mode represents the evolution of disks' rotations \cite{rot_mode3} and does not exist in frictionless disks.
In addition, Eq.\ [\ref{eq:motion}] describes purely harmonic oscillations of the disks around initial positions
such that any anharmonic behavior,\ e.g.\ due to opening and closing contacts \cite{saitoh10}, is not taken into account.
%
\begin{table}
\centering
\caption{The L, TR, and RT modes excited by different combinations of the amplitudes,
$\mathbf{A}=(A_x,A_y)$ and $A_\theta$, and the wave vector, $\mathbf{k}=(k_x,k_y)$,
where $A_x$ and $A_y$ are scaled by $d_0/t_0$ with the mean disk diameter, $d_0$, and $A_\theta$ is scaled by $t_0^{-1}$.}
\label{tab:modes}
\begin{tabular}{lccccc}
& $A_x$ & $A_y$ & $A_\theta$ & $k_x$ & $k_y$ \\
\midrule
L  & $10^{-3}$ & $0$ & $0$ & $k$ & $0$ \\
   & $0$ & $10^{-3}$ & $0$ & $0$ & $k$ \\
TR & $10^{-3}$ & $0$ & $0$ & $0$ & $k$ \\
   & $0$ & $10^{-3}$ & $0$ & $k$ & $0$ \\
RT & $0$ & $0$ & $10^{-3}$ & $k$ & $0$ \\
   & $0$ & $0$ & $10^{-3}$ & $0$ & $k$ \\
\bottomrule
\end{tabular}
\end{table}

\begin{figure}
\centering
\includegraphics[width=.8\linewidth]{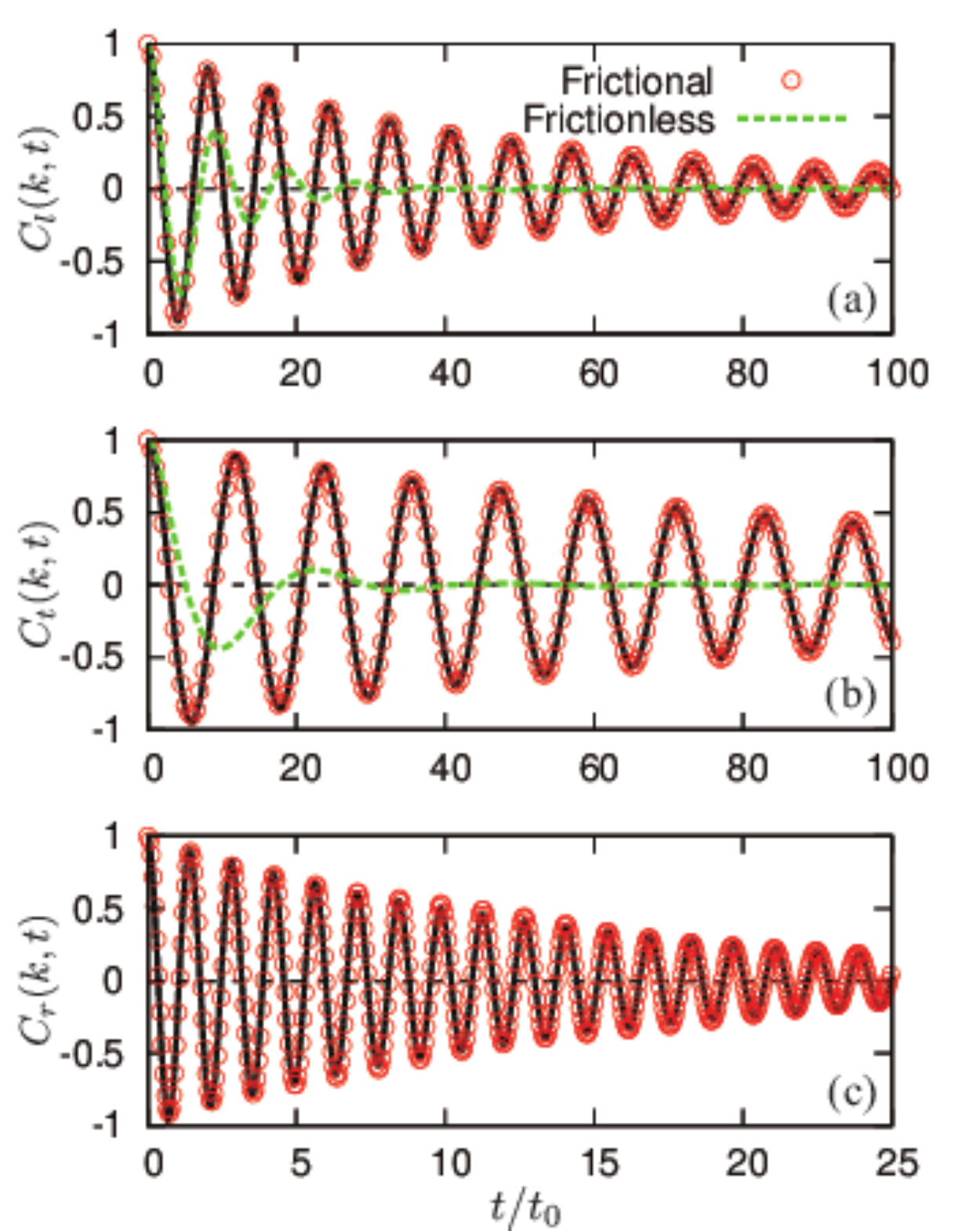}
\caption{
The VAFs of (a) longitudinal, (b) transverse, and (c) rotational velocities, where we used $\rho=1$ and $kd_0\simeq0.38\pi$.
The open circles are obtained from numerical solutions of Eq.\ [\ref{eq:motion}] and the solid lines represent damped oscillations,\ Eq.\ [\ref{eq:damped}].
In (a) and (b), our results of frictionless disks ($\rho=0$) are also shown (green dotted lines).
Note the different horizontal scale in (c).}
\label{fig:auto}
\end{figure}

From numerical solutions of Eq.\ [\ref{eq:motion}], we calculate normalized velocity autocorrelation functions (VAFs)
of longitudinal, transverse, and rotational velocities as $C_l(k,t)$, $C_t(k,t)$, and $C_r(k,t)$, respectively, where $k\equiv|\mathbf{k}|$ is the wave number.
Figure \ref{fig:auto} shows the time development of the VAFs (open circles), where we also plot our results of frictionless disks (green dotted lines in (a) and (b)).
As expected, in both frictional and frictionless disk systems, the oscillations of the L mode are faster than those of the TR mode (Figs.\ \ref{fig:auto}(a) and (b)).
In addition, the oscillations of the L and TR modes become faster in frictional disk systems, implying that the tangential forces increase the macroscopic elastic constants.
We also note that the decay of the VAFs, which is caused by \emph{scattering attenuation} of the L and TR modes, becomes weaker in frictional disks.
On the other hand, the oscillation of the RT mode is much faster than those of the L and TR modes (Fig.\ \ref{fig:auto}(c)),
indicating that eigen modes in the high frequency-band (Fig.\ \ref{fig:vdof}) are closely related to micropolar rotations of frictional disks.
Moreover, the VAF of the RT mode decays much faster, implying a stronger scattering attenuation.
%

\begin{figure*}
\centering
\includegraphics[width=17.8cm]{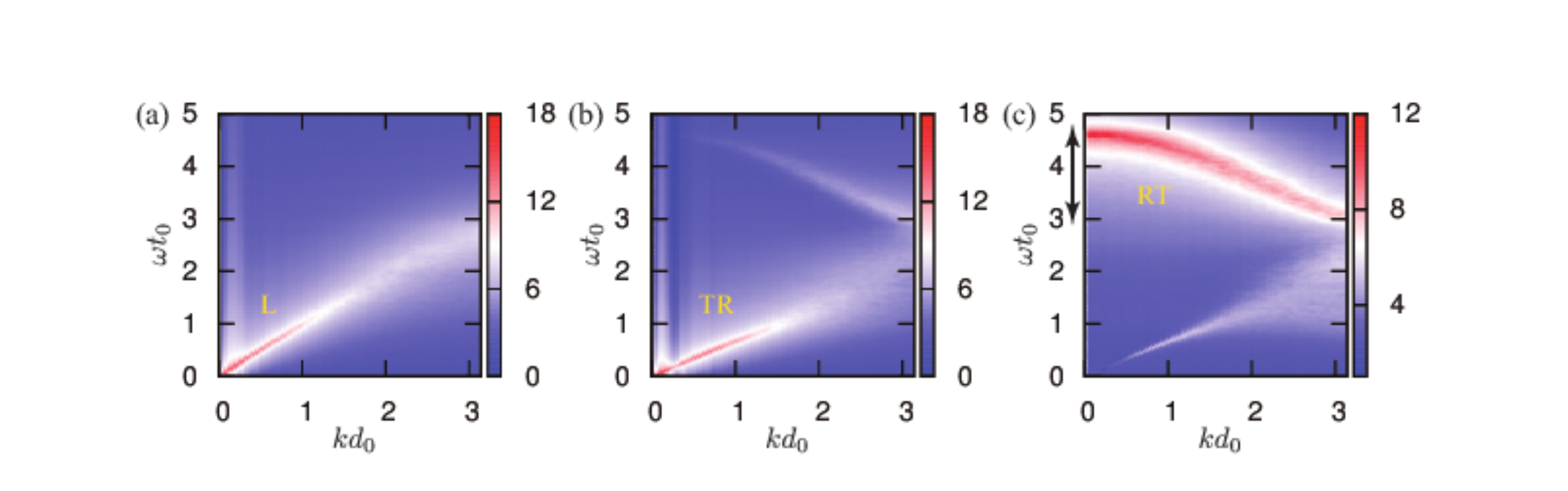}
\caption{
Dispersion relations of (a) the longitudinal (L), (b) transverse (TR), and (c) rotational (RT) modes, where the stiffness ratio is given by $\rho=1$.
The gray scale represents logarithms of power spectra,\ i.e.\ (a) $\log S_l(\omega,k)$, (b) $\log S_t(\omega,k)$, and (c) $\log S_r(\omega,k)$,
where the frequency and wave number are scaled by using $t_0$ and $d_0$, respectively.
The frequency-band of the RT mode is indicated by the double-headed vertical arrow in (c).}
\label{fig:spec}
\end{figure*}

To further study the evolution of elastic waves, we calculate power spectra of longitudinal, transverse, and rotational velocities
as $S_l(k,\omega)$, $S_t(k,\omega)$, and $S_r(k,\omega)$, respectively, for various wave number, $k$, in Eq.\ [\ref{eq:initial}].
Figure \ref{fig:spec} displays logarithms of the power spectra,\ i.e.\ $\log S_\alpha(k,\omega)$ ($\alpha=l,t,r$), which show dispersion relations of the elastic waves.
In the dispersion relations of both the L and TR modes (Figs.\ \ref{fig:spec}(a) and (b)), we observe strong ordinary acoustic branches,
where the speed defined as the slope, $\lim_{k\rightarrow0}\omega/k$, of the L mode is larger than that of the TR mode.
On the other hand, the RT mode exhibits an optical-like branch (Fig.\ \ref{fig:spec}(c)) which exists only in a high frequency regime ($2.5\le\omega t_0\le4.7$) as indicated by the double-headed vertical arrow in Fig.\ \ref{fig:spec}(c).
Because the vDOS is given by an integral of the dispersion relation over the whole wave number range \cite{psheng}, the high frequency-band in the vDOS (Fig.\ \ref{fig:vdof}) is the result of the optical-like branch.
In addition, the weak optical-like branch in Fig.\ \ref{fig:spec}(b) and the weak acoustic branch in Fig.\ \ref{fig:spec}(c) mean that rotations are always coupled with transverse (shear) motions \cite{rot_mode3}.
%

\subsection*{Lattice-based model}
The optical-like branch of the RT mode, which is spanning the high frequency-band, is a striking feature of frictional disk systems.
To quantitatively analyze its properties, we extract dispersion relations from numerical solutions of Eq.\ [\ref{eq:motion}].
For this purpose, we fit damped oscillations to the VAFs as
\begin{equation}
C_\alpha(k,t) = e^{-\gamma_\alpha(k)t}\cos\omega_\alpha(k)t
\label{eq:damped}
\end{equation}
($\alpha=l,t,r$), where the dispersion relation of each mode is given by the dominant frequency, $\omega_\alpha(k)$,
and the scattering attenuation of each mode is quantified by the \emph{attenuation coefficient}, $\gamma_\alpha(k)$
\footnote{The Fourier transform of Eq.\ [\ref{eq:damped}] is the Lorentzian, $S_\alpha(k,\omega)=\gamma_\alpha(k)/[\{\omega-\omega_\alpha(k)\}^2+\gamma_\alpha(k)^2]$,
which also well describes the power spectra (Fig.\ \ref{fig:spec}) except for the weak optical-like and acoustic branches in Figs.\ \ref{fig:spec}(b) and (c), respectively.} \cite{sound0}.
The solid lines in Fig.\ \ref{fig:auto} represent the damped oscillations,\ Eq.\ [\ref{eq:damped}], where we confirm perfect agreements with the VAFs by adjusting the parameters, $\omega_\alpha(k)$ and $\gamma_\alpha(k)$.
In SI, we summarize our results of the L and TR modes:
Small dips are observed in the phase speeds, $c_\alpha(k)\equiv\omega_\alpha(k)/k$ (Fig.\ S2),
while the attenuation coefficients obey the Rayleigh prediction of scattering attenuation,\ i.e.\ $\gamma_\alpha(k)\sim k^3$ ($\alpha=l,t$), which is not the case for frictionless disks (Fig.\ S3).

\begin{figure}
\centering
\includegraphics[width=.8\linewidth]{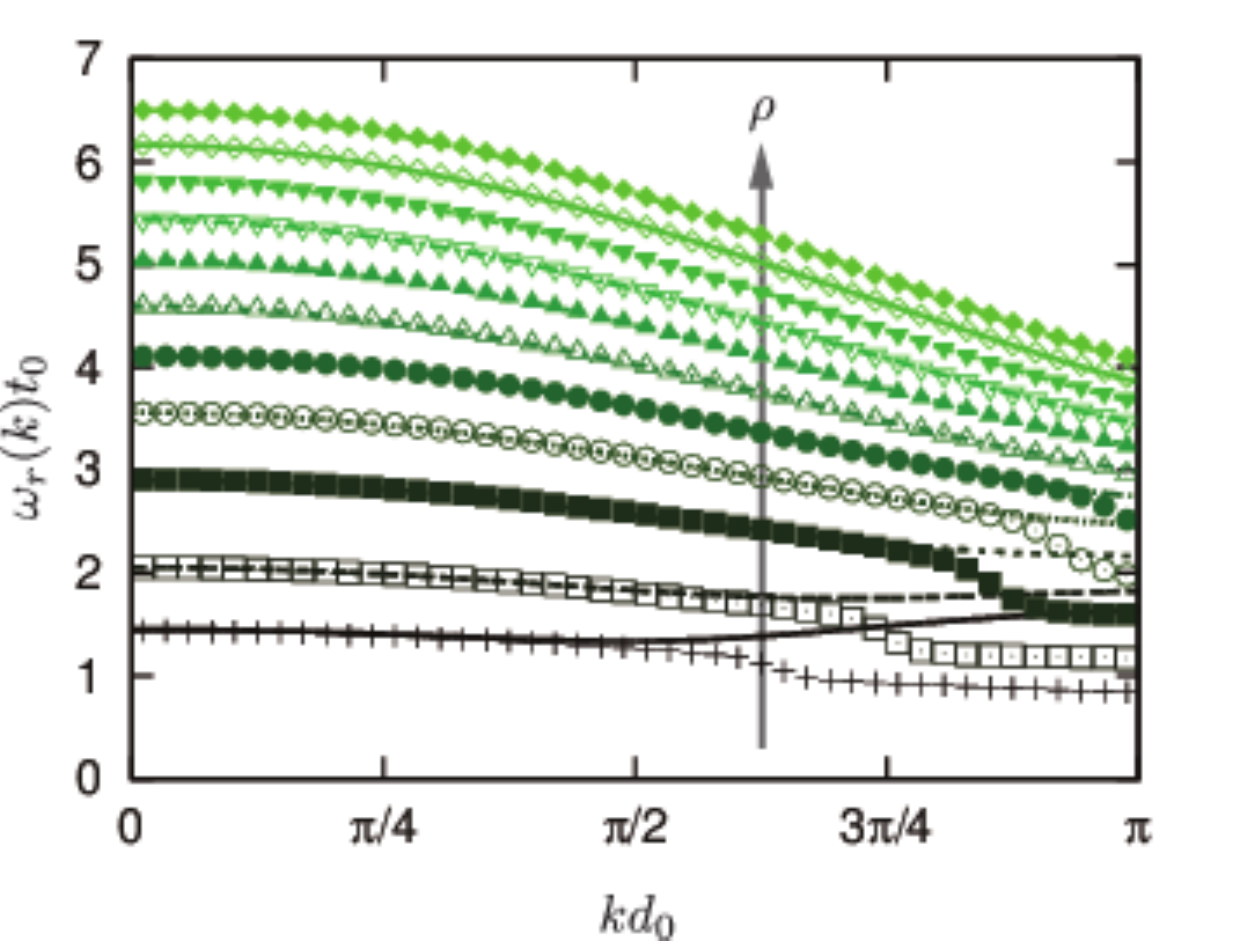}
\caption{
Dispersion relations of the RT mode, where the symbols are $\omega_r(k)$ in Eq.\ [\ref{eq:damped}] and lines are the model predictions,\ Eq.\ [\ref{eq:omega_r(k)}].
The stiffness ratio increases as $\rho=0.1$, $0.2$, $0.4$, $0.6$, $0.8$, $1.0$, $1.2$, $1.4$, $1.6$, $1.8$, and $2.0$ (as indicated by the arrow).}
\label{fig:dispersion_RT}
\end{figure}

Figure \ref{fig:dispersion_RT} displays dispersion relations of the optical-like RT modes, $\omega_r(k)$ (symbols),
where the frequency-band shifts to higher frequencies with increasing the stiffness ratio, $\rho$ (as indicated by the arrow).
To explain such a dependence of optical-like branches on the stiffness ratio, we modify the discrete model of frictional grains on a lattice \cite{rot_mode3},
where the dispersion relations of the TR and RT modes are given by
\begin{eqnarray}
\omega_t(k) &=& a_t\sqrt{f(k,\rho)-\sqrt{g(k,\rho)}}~,\label{eq:omega_t(k)}\\
\omega_r(k) &=& a_r\sqrt{f(k,\rho)+\sqrt{g(k,\rho)}}~,\label{eq:omega_r(k)}
\end{eqnarray}
respectively.
In Eqs.\ [\ref{eq:omega_t(k)}] and [\ref{eq:omega_r(k)}], the prefactor, $a_\alpha$ ($\alpha=t,r$), is introduced to represent the decrease of sound speed by dispersion (due to disorder) and the functions are given by
\begin{eqnarray}
f(k,\rho) &=& 2\sin^2(kl)+9\rho\cos^2(kl)+11\rho~, \label{eq:f(k)}\\
g(k,\rho) &=& 4\sin^4(kl)+\rho(300\rho-4)\cos^2(kl) \nonumber\\
& & -\rho\left(\frac{121}{4}\rho+21\right)\sin^2(2kl)+\rho(\rho+4)~. \label{eq:g(k)}
\end{eqnarray}
In Eqs.\ [\ref{eq:f(k)}] and [\ref{eq:g(k)}], we estimate the length scale, $l$, by considering the first Brillouin zone, $|kl|\le\pi/2$:
Equating the maximum wave number, $k_\mathrm{max}\equiv\pi/2l$, in the model with $\pi/d_0$, we obtain $l\approx d_0/2$, where $d_0$ is the mean disk diameter.
The lines in Fig.\ \ref{fig:dispersion_RT} are the model predictions of $\omega_r(k)$ (Eq.\ [\ref{eq:omega_r(k)}]), where we used $a_r\simeq0.73t_0^{-1}$ and $l\simeq0.446d_0$ for all $\rho$.
In this figure, all the dispersion relations are consistent with the lattice model for long wave lengths, $kd_0\lesssim\pi/2$,
as the difference in microstructures,\ i.e.\ order and disorder, should not affect the long wave properties (see Figs.\ S5 and S6 for the L and TR modes).
In addition, if the stiffness ratio is fairly large, $\rho>1$, the dispersion relations are perfectly described by the model for all $k$.
However, if the stiffness ratio is small, $\rho<1$, we observe deviations from the model for short wave lengths, $k>k_c$,
where the characteristic wave number, $k_c$, monotonously decreases with decreasing stiffness ratio.
%

\subsection*{Transition behavior of the RT mode}
\begin{figure}
\centering
\includegraphics[width=\linewidth]{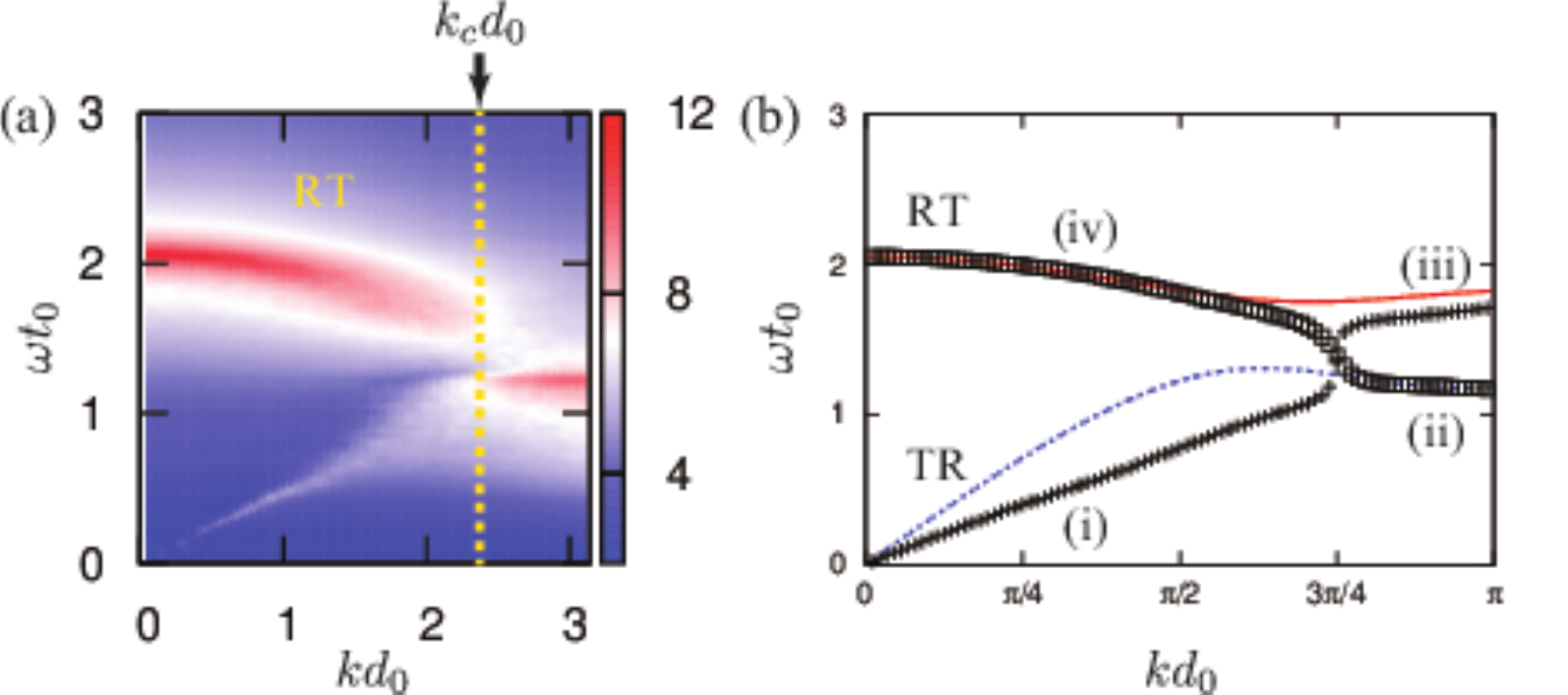}
\caption{
(a) A three dimensional plot of $\log S_r(\omega,k)$, where the characteristic wave number, $k_c$, is indicated by the arrow.
(b) Dispersion relations obtained from the fitting to VAFs,\ Eq.\ [\ref{eq:damped}], where the crosses and open squares are $\omega_t(k)$ and $\omega_r(k)$, respectively.
The dotted and solid lines represent the models,\ Eqs.\ [\ref{eq:omega_t(k)}] and [\ref{eq:omega_r(k)}], respectively.
In both (a) and (b), the stiffness ratio is $\rho=0.2$.}
\label{fig:switching}
\end{figure}

To clarify the deviations in the short wave lengths, $k>k_c$, we closely look at the dispersion relations with small stiffness ratios, $\rho<1$.
Figure \ref{fig:switching} shows (a) the logarithm of the power spectrum, $S_r(k,\omega)$,
and (b) dispersion relations, $\omega_t(k)$ and $\omega_r(k)$, obtained from the fitting to VAFs,\ Eq.\ [\ref{eq:damped}], where the stiffness ratio is given by $\rho=0.2$.
In Fig.\ \ref{fig:switching}(a), the optical-like branch ends at $k=k_c$ (as indicated by the arrow) and drops to another branch in $k>k_c$, which makes a small gap in the frequency between $1.2<\omega t_0<1.5$.
Accordingly, in Fig.\ \ref{fig:switching}(b), the dispersion relation of the RT mode (open squares) suddenly drops to lower values at $k=k_c$ (from the branch (iv) to (ii)),
while that of the TR mode (pluses) jumps from the branch (i) to (iii).
Thus, the RT (TR) mode dominates lower (higher) frequencies in the short wave lengths.
The dispersion relation of the RT mode in $k<k_c$ agrees with the model of $\omega_r(k)$,\ Eq.\ [\ref{eq:omega_r(k)}] (the solid line in Fig.\ \ref{fig:switching}(b)),
while it is well explained by the model of $\omega_t(k)$,\ Eq.\ [\ref{eq:omega_t(k)}] (the dotted line in Fig.\ \ref{fig:switching}(b)), if $k>k_c$.
Therefore, it seems that disks' rotations exhibit a transition from the optical-like fast oscillations to the acoustic wave behavior.
%

\begin{figure}
\centering
\includegraphics[width=\linewidth]{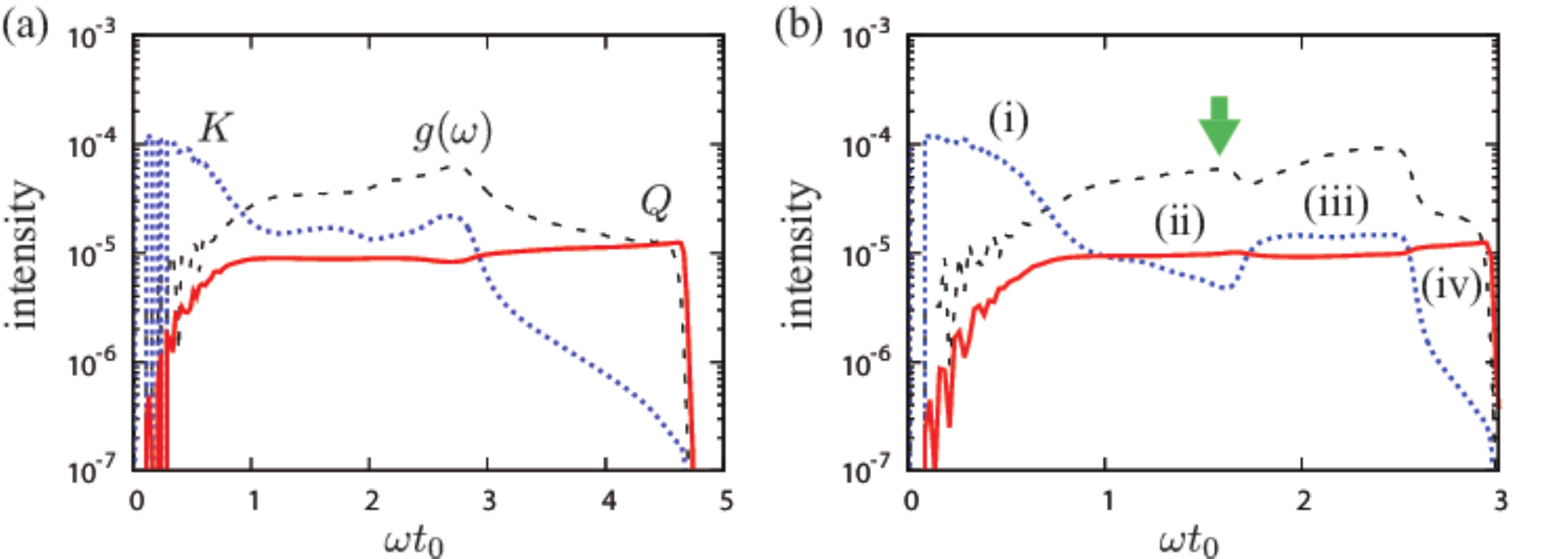}
\caption{
Semi-logarithmic plots of intensities of $K$ (blue dotted lines), $Q$ (red solid lines), and $g(\omega)$ multiplied by $5\times10^{-3}$ (dashed lines),
where the stiffness ratios are given by (a) $\rho=1$ and (b) $\rho=0.4$.
The green vertical arrow in (b) indicates the secondary peak of vDOS (as in Fig.\ \ref{fig:vdof}).}
\label{fig:kene}
\end{figure}

To further examine the transition behavior, we study the eigen vectors of the dynamical matrix, $\bar{\mathbf{u}}_q=\{\bar{u}_{ix},\bar{u}_{iy},\bar{\theta}_i\}$ (i.e.\ displacements associated with each eigen frequency).
Here, we quantify kinetic energy for each degree of freedom by $K_\nu\equiv\sum_i m\dot{\bar{u}}_{i\nu}^2/2N$ ($\nu=x,y$) and $Q\equiv\sum_i I_i\dot{\bar{\theta}}_i^2/2N$, where $\dot{\bar{\mathbf{u}}}_q\equiv\bar{\mathbf{u}}_q/t_0$.
The translational energy, $K\equiv(K_x+K_y)/2$, and rotational one, $Q$, represent the intensity of the L and TR modes and that of the RT mode, respectively.
As shown in Fig.\ \ref{fig:kene}(a), if the stiffness ratio is large enough, $K$ and $Q$ dominate low ($\omega t_0\lesssim 1$) and high ($\omega t_0\gtrsim 3$) eigen frequencies, respectively.
In this case, the acoustic L and TR branches at low frequencies are well separated from the optical-like RT branch at high frequencies (as in Fig.\ \ref{fig:spec}),
where the shoulder of the vDOS (the dashed line in Fig.\ \ref{fig:kene}(a)) for high frequencies, $3<\omega t_0<4.6$, is mostly owned by disks' rotations, $Q$.
However, if the stiffness ratio is small (Fig.\ \ref{fig:kene}(b)), there are four regions,\ i.e.\ (i) $K>Q$, (ii) $K<Q$, (iii) $K>Q$, and (iv) $K<Q$, corresponding to the branches (i)-(iv) in Fig.\ \ref{fig:switching}(b).
Therefore, the transition from the optical-like to acoustic behavior of disks' rotations is also evidenced by the eigen vectors (see also Fig.\ S6 and S8).
Because the vDOS is given by integrating dispersion relations over the whole wave number range,
the secondary peak (indicated by the vertical arrow in Fig.\ \ref{fig:kene}(b)) represents the upper end of (ii),\ i.e.\ the RT mode on the acoustic branch.
%

\subsection*{Frequency-band of the RT mode}
\begin{figure}
\centering
\includegraphics[width=.8\linewidth]{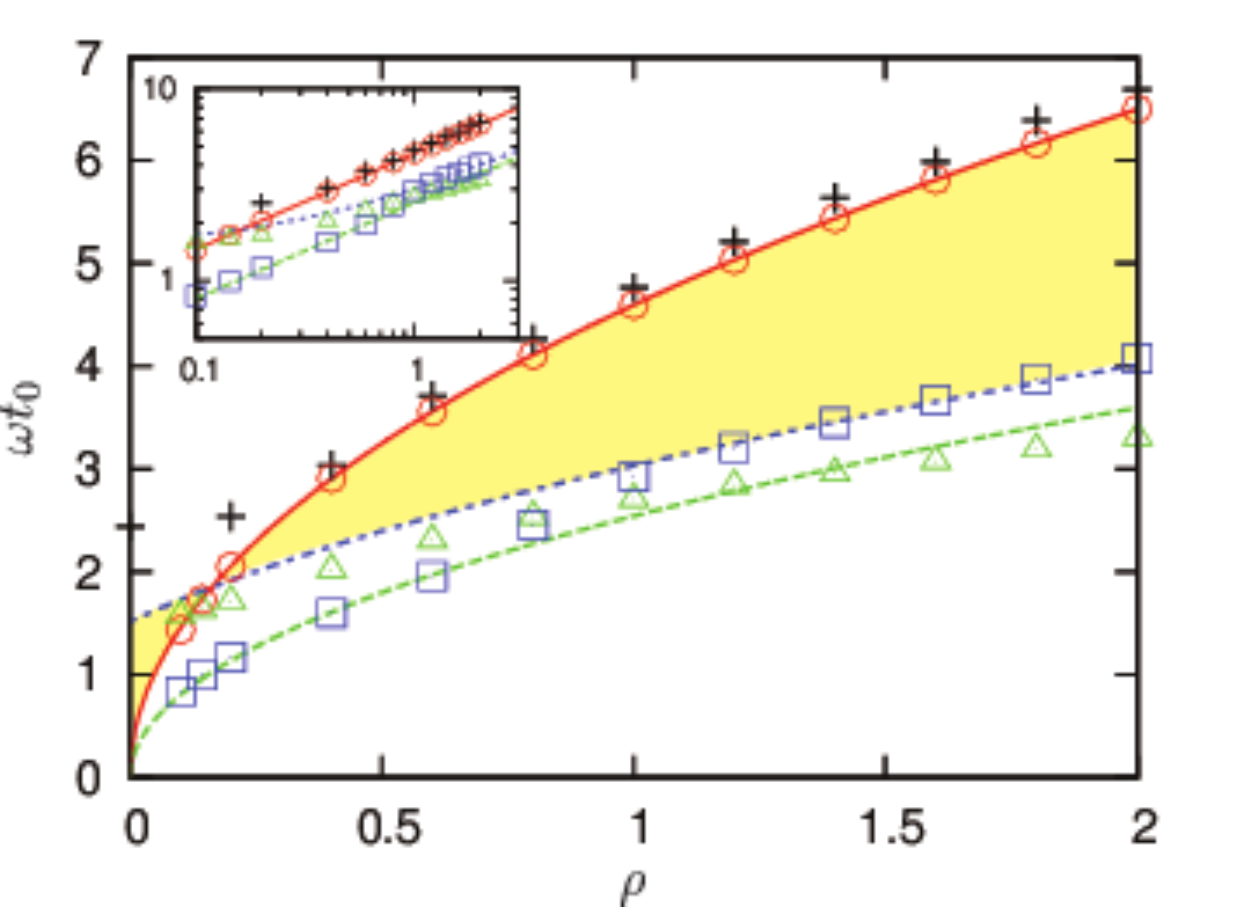}
\caption{
Frequency-bands of the RT mode, where the red solid, blue dotted, and green dashed lines are the model predictions of $\omega_r(0)$, $\omega_r(k_\mathrm{max})$, and $\omega_t(k_\mathrm{max})$, respectively.
The shaded region is the model prediction of the frequency-bands, $\omega_r(0)<\omega<\omega_r(k_\mathrm{max})$ for $\rho<1/7$ and $\omega_r(k_\mathrm{max})<\omega<\omega_r(0)$ otherwise.
The open circles and squares are numerical results of limit frequencies, $\omega_r(0)$ and $\omega_r(k_\mathrm{max})$, respectively.
The crosses are the cutoff frequencies, $\omega_c$, estimated from vDOS (Fig.\ \ref{fig:vdof}), while the open triangles are numerical results of the limit, $\omega_t(k_\mathrm{max})$.
The inset shows double logarithmic plots, where the red solid and green dashed lines have the slope, $1/2$.
\label{fig:band_RT}}
\end{figure}

The lattice-based model (Eq.\ [\ref{eq:omega_r(k)}]) also predicts frequency-bands of the RT mode \cite{rot_mode3},\
i.e.\ $\omega_r(k_\mathrm{max})\le\omega\le\omega_r(0)$ for $\rho>1/7$ and $\omega_r(0)\le\omega\le\omega_r(k_\mathrm{max})$ otherwise,
where the frequency-band limits coincide,\ i.e.\ $\omega_r(0)=\omega_r(k_\mathrm{max})$, at $\rho=1/7$.
The shaded region in Fig.\ \ref{fig:band_RT} is the model prediction of the frequency-bands,
where the solid, dotted, and dashed lines represent $\omega_r(0)=t_0^{-1}c_r\sqrt{40\rho}$, $\omega_r(k_\mathrm{max})=2t_0^{-1}c_r\sqrt{3\rho+1}$, and $\omega_t(k_\mathrm{max})=t_0^{-1}c_t\sqrt{10\rho}$, respectively.
Our numerical results of the limit, $\omega_r(0)$ (open circles), agree well with the model (solid line) over the whole range of stiffness ratios.
In addition, the cutoff frequency of vDOS, $\omega_c$ (crosses), shows qualitatively the same behavior as $\omega_r(0)$, where $\omega_c$ is slightly higher because of the non-sharp upper limit of the dispersion relation.
If the stiffness ratio is large, $\rho>1$, numerical results of the limit, $\omega_r(k_\mathrm{max})$ (open squares), are explained by the model (dotted line).
However, if the stiffness ratio is small, $\rho<1$, the limit switches to the acoustic branch (dashed line), resulting from the transition of the RT mode (Fig.\ \ref{fig:switching}).
Therefore, different from granular crystals \cite{rot_mode3}, the upper and lower limits of the RT mode, $\omega_r(0)$ and $\omega_r(k_\mathrm{max})$, in disordered frictional disks are not reversed.
We also note that the limit frequency of the TR mode, $\omega_t(k_\mathrm{max})$ (open triangles), slightly accedes to the model predictions.
%

\section*{Discussion}
In this study, we have numerically investigated sound in frictional disk systems with the focus on (i) the configurational disorder and (ii) the strength of tangential forces.
Employing the dynamical matrix of disordered frictional disks, we first analyze their eigen frequencies.
We find that a high frequency-band or shoulder in the vDOS, which is characteristic of frictional disk systems, develops for higher frequencies with increasing the stiffness ratio (or tangential forces).
Numerically solving the equations of motion, we then study how longitudinal, transverse, and rotational elastic standing waves evolve in the system.
Even strong tangential coupling, we observe that the rotational mode exhibits much faster oscillations than those of the acoustic longitudinal and transverse modes.
The fast oscillations of micropolar rotations are represented by an optical-like branch in the dispersion relation of the rotational mode.
We explain the characteristic optical-like dispersion relations by introducing a lattice-based model.
The lattice-based model perfectly describes our numerical results in the case that tangential forces are strong enough and in any cases for small wave numbers.
It is remarkable, however, that the rotational mode exhibits a transition to the acoustic branch in the short wave length regime if the tangential forces are comparable with or smaller than the normal forces.
The transition from the optical-like to acoustic branches is also evidenced by our analysis of eigen-vectors.
We find that the rotational mode in the acoustic branch is also well described by the lattice-based model.

We conclude that rotational sound exhibits a characteristic optical-like branch in the dispersion relation even in disordered systems.
If the tangential forces are comparable with or smaller than the normal forces, it changes to the acoustic branch at a characteristic wave length
so that configurational disorder enables the acoustic behavior of micropolar rotations at small enough scales.

In our numerical simulations, we model both the normal and tangential forces by elastic springs.
Thus, the standing waves presented are purely elastic so that total energy is conserved throughout simulations, where the decay of VAFs is solely caused by scattering attenuation (and not by energy dissipation).
However, in reality, dissipative forces,\ e.g.\ the Coulomb or sliding friction and viscous forces, also exist between the disks in contact.
To take into account such dissipation of energy, we need some generalizations of our model as in \cite{fri_vm0,vis_vm0}.
Therefore, effects of these dissipative forces on our results are left to future work.
Similarly, it is interesting to study how other interaction forces,\ e.g.\ the rolling resistance \cite{rot_mode3}
and the attractive interaction due to capillary bridges in wet granular material \cite{sound_prop2,stefan-N7a,stefan-N7b}, affect the results.
Moreover, the influence of microstructure \cite{psheng},\ e.g.\ size-distributions and polydispersity, on the rotational sound requires more research.
For practical purposes, numerical studies in three dimensions are crucial,
where an additional degree of freedom,\ i.e.\ the twisting motion of frictional spheres in contact, enables a \emph{pure rotational} (R) \emph{mode} \cite{rot_mode1,rot_mode3}.
Therefore, further studies are needed to clarify how configurational disorder affects the L, TR, RT, and R modes in three-dimensional granular media.
In addition, not only dispersion and scattering attenuation, but also wave diffusion \cite{sound_diff1}
and localization phenomena \cite{sound_local1,sound_local2} are important aspects of sound in granular material.
%
\matmethods{
\subsection*{Mass matrix and elastic energy}
In the equations of motion,\ Eq.\ [\ref{eq:motion}], the $3N\times3N$ mass matrix is given by
\begin{equation}
\mathcal{M}=
\begin{pmatrix}
	m_i & 0 & 0 \\
	0 & m_i & 0 \\
	0 & 0 & I_i
\end{pmatrix}_{i=1,\dots,N}~,
\label{eq:mm}
\end{equation}
where $m_i$ and $I_i=m_id_i^2/8$ with the disk diameter, $d_i$, are the mass and moment of inertia of the disk $i$, respectively.

We introduce elastic energy of disordered frictional disks as the sum of pairwise potentials,\ i.e.\ $E=\sum_{i>j}e_{ij}$.
The pairwise potential is decomposed into harmonic potentials stored in normal and tangential directions as
\begin{equation}
e_{ij}=\frac{k_n}{2}\xi_{ij}^2+\frac{k_t}{2}\mathbf{u}_{ij}^{\perp 2}~,
\label{eq:eij}
\end{equation}
where $k_n$ and $k_t$ are the normal and tangential stiffness, respectively.
In Eq.\ [\ref{eq:eij}], $\xi_{ij}\equiv(d_i+d_j)/2-r_{ij}>0$ represents the overlap between the disks ($i$ and $j$) in contact,
where $r_{ij}\equiv|\mathbf{r}_{ij}|$ with the relative position between the disks, $\mathbf{r}_{ij}\equiv\mathbf{r}_i-\mathbf{r}_j$, is the inter-particle distance.
In addition, $\mathbf{u}_{ij}^\perp\equiv\mathbf{u}_{ij}-\mathbf{u}_{ij}^\parallel-\boldsymbol{\theta}_{ij}\times\mathbf{n}_{ij}$ is the relative displacement in tangential direction,
where we introduced relative displacements as $\mathbf{u}_{ij}\equiv\mathbf{u}_i-\mathbf{u}_j$, $\mathbf{u}_{ij}^\parallel\equiv(\mathbf{u}_{ij}\cdot\mathbf{n}_{ij})\mathbf{n}_{ij}$,
and $\boldsymbol{\theta}_{ij}\equiv(d_i\theta_i+d_j\theta_j)\mathbf{n}_z/2$ with the normal unit vector, $\mathbf{n}_{ij}\equiv\mathbf{r}_{ij}/r_{ij}$, and out of $xy$-plane unit vector, $\mathbf{n}_z$ (parallel to the $z$-axis).
\subsection*{Disordered configurations}
We generate initial disordered configurations by MD simulations.
To avoid crystallization, we randomly distribute a $50:50$ binary mixture of $N$ frictionless disks ($k_t=0$) in a periodic box,
where different kinds of disks have the same mass, $m$, and different diameters, $d_L$ and $d_S$ (with ratio $d_L/d_S=1.4$).
Then, we minimize elastic energy, $E_n=\sum_{i>j}k_n\xi_{ij}^2/2$, with the aid of the FIRE algorithm \cite{FIRE}
and stop the energy minimization once the maximum of disk accelerations becomes less than $10^{-6}k_nd_0/m$, where $d_0\equiv(d_L+d_S)/2$ is the mean disk diameter.
After the energy minimization, no potential energy is stored in tangential direction so that the system is still in mechanical equilibrium even if we switch on the tangential forces,\ i.e.\ $E=E_n$ even though $k_t>0$.
Therefore, our systems are initially \emph{unstressed} in tangential direction \cite{saitoh13}.
Note that \emph{stressed systems} can be made if we prepare initial disordered configurations with tangential forces ($k_t>0$) which make the configurations history-dependent.
\subsection*{The VAFs and power spectra}
From numerical solutions of Eq.\ [\ref{eq:motion}],\ i.e.\ $\{\dot{\mathbf{u}}_i(t),\dot{\theta}_i(t)\}$ ($i=1,\dots,N$),
we calculate Fourier transforms of disks' velocities as
\begin{equation}
\{\dot{\mathbf{u}}_\mathbf{k}(t),\dot{\theta}_\mathbf{k}(t)\}=\sum_{i=1}^N\{\dot{\mathbf{u}}_i(t),\dot{\theta}_i(t)\}e^{-I\mathbf{k}\cdot\mathbf{r}_i(t)}~,
\end{equation}
where the position, $\mathbf{r}_i(t)$, is also given by numerical integrations of Eq.\ [\ref{eq:motion}].
The longitudinal and transverse velocities are defined as
\begin{eqnarray}
\dot{\mathbf{u}}_\mathbf{k}^\parallel(t) &\equiv& \{\dot{\mathbf{u}}_\mathbf{k}(t)\cdot\hat{\mathbf{k}}\}\hat{\mathbf{k}}~,\\
\dot{\mathbf{u}}_\mathbf{k}^\perp(t) &\equiv& \dot{\mathbf{u}}_\mathbf{k}(t)-\dot{\mathbf{u}}_\mathbf{k}^\parallel(t)~,
\end{eqnarray}
respectively, where $\hat{\mathbf{k}}\equiv\mathbf{k}/k$ with $k\equiv|\mathbf{k}|$ is a unit vector parallel to the wave vector.
Then, the normalized VAFs of longitudinal, transverse, and rotational velocities are given by
$C_l(k,t)=\langle\dot{\mathbf{u}}_\mathbf{k}^\parallel(t)\cdot\dot{\mathbf{u}}_{-\mathbf{k}}^\parallel(0)\rangle/\langle|\dot{\mathbf{u}}_\mathbf{k}^\parallel(0)|^2\rangle$,
$C_t(k,t)=\langle\dot{\mathbf{u}}_\mathbf{k}^\perp(t)\cdot\dot{\mathbf{u}}_{-\mathbf{k}}^\perp(0)\rangle/\langle|\dot{\mathbf{u}}_\mathbf{k}^\perp(0)|^2\rangle$,
and $C_r(k,t)=\langle\dot{\theta}_\mathbf{k}(t)\dot{\theta}_{-\mathbf{k}}(0)\rangle/\langle|\dot{\theta}_\mathbf{k}(0)|^2\rangle$, respectively.

The power spectra of longitudinal, transverse, and rotational velocities are introduced as
$S_l(k,\omega)=\langle|\tilde{\dot{\mathbf{u}}}_\mathbf{k}^\parallel(\omega)|^2\rangle$,
$S_t(k,\omega)=\langle|\tilde{\dot{\mathbf{u}}}_\mathbf{k}^\perp(\omega)|^2\rangle$,
and $S_r(k,\omega)=\langle|\tilde{\dot{\theta}}_\mathbf{k}(\omega)|^2\rangle$, respectively,
where the Fourier transforms are given by
$\tilde{\dot{\mathbf{u}}}_\mathbf{k}^\parallel(\omega)\equiv\int_0^\infty\dot{\mathbf{u}}_\mathbf{k}^\parallel(t)e^{I\omega t}dt$,
$\tilde{\dot{\mathbf{u}}}_\mathbf{k}^\perp(\omega)\equiv\int_0^\infty\dot{\mathbf{u}}_\mathbf{k}^\perp(t)e^{I\omega t}dt$,
and $\tilde{\dot{\theta}}_\mathbf{k}(\omega)\equiv\int_0^\infty\dot{\theta}_\mathbf{k}(t)e^{I\omega t}dt$.
} 
\showmatmethods{} 
\acknow{We thank A. Merkel, X. Jia, H. Mizuno, B.P. Tighe, V. Magnanimo, and H. Cheng for fruitful discussions.
This work was financially supported by Kawai Foundation for Sound Technology \& Music and KAKENHI Grant No.\ 16H04025 and No.\ 18K13464 from JSPS.
A part of computation has been carried out at the Yukawa Institute Computer Facility.}
\showacknow{}
\bibliography{sound}
\end{document}